\documentclass[aip,apl,reprint]{revtex4-1}
\usepackage[dvips]{graphicx}
\usepackage{color}
\usepackage{amsmath}
\usepackage{comment}

\begin{document}
\title{Effects of frequency mixing on Shapiro-step formations in sliding charge-density-waves} 

\author{Yu Funami}
\email[]{funami@neq.ess.sci.osaka-u.ac.jp}
\author{Kazushi Aoyama}
\email[]{aoyama@ess.sci.osaka-u.ac.jp}
\affiliation{Department of Earth and Space Science, Graduate School of Science, Osaka University, Osaka 560-0043, Japan}


\date{\today}

\begin{abstract}
A one-dimensional charge density wave (CDW) is driven to slide by a dc electric field, carrying an electric current. In an additional ac field with frequency ${\omega}_{\mathrm{ex}}$, it is known that the sliding CDW can be synchronized to $\omega_{\mathrm{ex}}$, leading to the occurrence of Shapiro steps in the $I$-$V$ characteristics. Motivated by a recent experiment where ac fields with two frequencies $\omega_{\mathrm{ex}}$ and $\omega_{\mathrm{ex}}^{\prime}$ are simultaneously applied, we theoretically investigate the effects of frequency mixing on the Shapiro-step formation. Based on the Fukuyama-Lee-Rice model, we show that in addition to the main steps induced by $\omega_{\mathrm{ex}}$, satellite steps characterized by $\omega_{\mathrm{ex}}^{\prime}$ emerge. It is also found that with increasing the ac-field strength for $\omega_{\mathrm{ex}}^{\prime}$, each step width first exhibits a damped oscillation as in the one-frequency case, and then, exhibits a non-monotonic behavior. The origin of these behaviors and the relevance to the associated experiment are also discussed.
\end{abstract}


\maketitle 

A charge density wave (CDW) typically emerging in quasi-one-dimensional conductors is a spatial order of electron density,\cite{Gruner1988,Monceau2012} and its dynamics leads to a diversity of phenomena such as nonlinear conduction and interference effects.\cite{Thorne1996, Zybtsev2024}
Understanding these phenomena is important not only from the fundamental point of view but also for potential applications to electronic devices.\cite{Balandin2021}
Since the CDW is an electron condensate, it is subject to pinning by charge disorders originating from impurities and lattice defects.\cite{Lee1974,Liu2021}
By applying an external dc electric field $E_{\rm dc}$, the CDW is depinned and begins a collective translational motion, the so-called CDW sliding.\cite{Frohlich1954,Monceau1976,Ong1977,Fleming1979,Thorne1987a}
Then, the current-voltage ($I$-$V$) characteristics exhibit a nonlinear behavior due to an additional electric current carried by the sliding CDW $I_{\rm CDW}$. Similar depinning phenomena can be widely seen in different contexts\cite{Reichhardt2017} such as frictions and superconducting vortices.\cite{Matsukawa1994, Maeda2005, Bhattacharya1993, Kaji2022}
One particularly interesting aspect of the CDW sliding is a synchronization phenomenon.
In the presence of an ac field in addition to the dc field, i.e., $E=E_{\rm dc}+E_{\rm ac}\sin(\omega_{\rm ex}t)$, a sliding mode $\omega_\phi$ is synchronized to the ac-field frequency $\omega_{\rm ex}$, leading to the emergence of plateau regions in the $I$-$V$ characteristics, known as the Shapiro steps.\cite{Brown1984,Thorne1987,Thorne1988,Zettl1984,Zybtsev2020,Richard1982,Coppersmith1986,Matsukawa_JJAP_1987,Bhattacharya1987,Middleton1992,Higgins1993,McCarten1994}
Each Shapiro step is characterized by the associated synchronization condition $\omega_\phi=(p/q)\omega_{\rm ex}$ with integers $p$ and $q$.

Recently, the effects of frequency mixing on the Shapiro-step formation were experimentally studied,\cite{Nikonov2021} where external fields with two frequencies (the ratio of a higher frequency $\omega_{\rm ex}$ to a lower frequency $\omega^{\prime}_{\rm ex}$ is $400:50$ MHz)
were applied to the quasi-one-dimensional CDW conductor $\mathrm{NbS}_3$. It has been reported that in addition to the main Shapiro steps determined by the condition $\omega_\phi = (p/q)\omega_{\rm ex}$, satellite Shapiro steps appear such that they sandwich the main steps. For example, for the main step of $\omega_\phi=(1/1)\omega_{\rm ex}$, the satellite steps appear at $\omega_\phi = \omega_{\rm ex}\pm\omega^{\prime}_{\rm ex}$. As a result, the whole structure of the $I$-$V$ characteristics is different from that in the one-frequency case, but the difference in the step-formation mechanism is not yet understood.

In this Letter, to gain an insight into the origin of the satellite steps, we theoretically investigate the frequency mixing effects on the Shapiro step formation.
We will show that due to an emergent multi-frequency synchronization, the main steps induced by $\omega_{\rm ex}$ are accompanied by satellite steps characterized by $\omega_{\rm ex}^\prime$ and that in contrast to the one-frequency case where the step width is a Bessel-type damped oscillation function of $E_{\rm ac}$\cite{Zettl1984,Thorne1987,McCarten1994,Zybtsev2020,Thorne1988} [see Figs. \ref{fig:IV-single}(b) and \ref{fig:IV-single}(c)], in the two-frequency case, the step width exhibits an oscillatory but undamped behavior (see Fig. \ref{fig:step-simulation}).

We start by introducing the one-dimensional CDW whose electron density is expressed as
\begin{equation}\label{eq:Charge-density}
\rho(x, t)=\rho_0+\rho_1 \cos [\phi(x, t)+Q x],
\end{equation}
where $\rho_0$ is the average electron density, and $\rho_1$ and $Q$ are the amplitude and wave number of the CDW modulation, respectively. The CDW dynamics can be described by the time evolution of the phase $\phi(x,t)$. When the CDW is driven by $E_{\rm dc}$, the CDW current $I_{\mathrm{CDW}}$ proportional to the sliding velocity $v=\frac{1}{Q}\frac{d\phi}{dt}$ flows,\cite{Gruner1988, Monceau2012,Frohlich1954,Ong1977,Monceau1976} showing a temporal oscillation with frequency $\omega_\phi$ called narrow band noise.\cite{Fleming1979,Thorne1987a,Bhattacharya1987,Richard1982,Thorne1988} Since the CDW frequency $\omega_{\mathrm{\phi}}$ is determined from the dispersion relation $\omega_{\mathrm{\phi}} = vQ =\frac{d \phi}{dt}$, the following relation holds:
\begin{equation}\label{CDW-current}
I_{\mathrm{CDW}}\propto \frac{d\phi}{dt} = \omega_{\mathrm{\phi}}.
\end{equation}
To discuss the Shapiro steps appearing in the $I$-$V$ characteristics, we need to analyze the time evolution of the CDW phase $\phi$.

Although several models for the CDW dynamics have been proposed,\cite{Thorne2005} here, we use the Fukuyama-Lee-Rice (FLR) model\cite{Fukuyama1976,Fukuyama1978,Lee1979} taking account of the pinning effect of randomly distributed impurities. 
Since the depinning phenomenon and low-frequency dynamics like the one observed in the two-frequency experiment\cite{Nikonov2021} can be described in the overdamped approximation,\cite{Zettl1982,Sridhar1985,Reagor1986,Sridhar1986} we use the equation of motion for the FLR model in the overdamped regime
\begin{equation}\label{eq:FLR-model}
\left(\gamma\frac{\partial}{\partial t}-v_{\mathrm{ph}}^2\nabla^2\right) \phi=P_{\mathrm{imp}} N_{\rm p}(x)\sin \left(\phi+Qx\right)+\frac{eQ}{m^*} E,
\end{equation}
where $e$ and $m^{\ast}$ are the electric charge and the effective mass of the CDW, respectively, $v_{\mathrm{ph}}$ is the phason velocity, and $\gamma$ is a phenomenologically introduced damping constant.
The pinning effect is incorporated in the nonlinear $\sin(\phi+Qx)$ term, where $P_{\mathrm{imp}}$ and $N_{\rm p}(x)=\sum_{i}\delta(x-R_i)$, respectively, denote a pinning strength and the distribution function of pinning sites whose positions are denoted by $R_i$. $P_{\mathrm{imp}}$ is assumed to be spatially uniform for simplicity.

\begin{figure}[t]
\begin{center}
\includegraphics[width=\columnwidth]{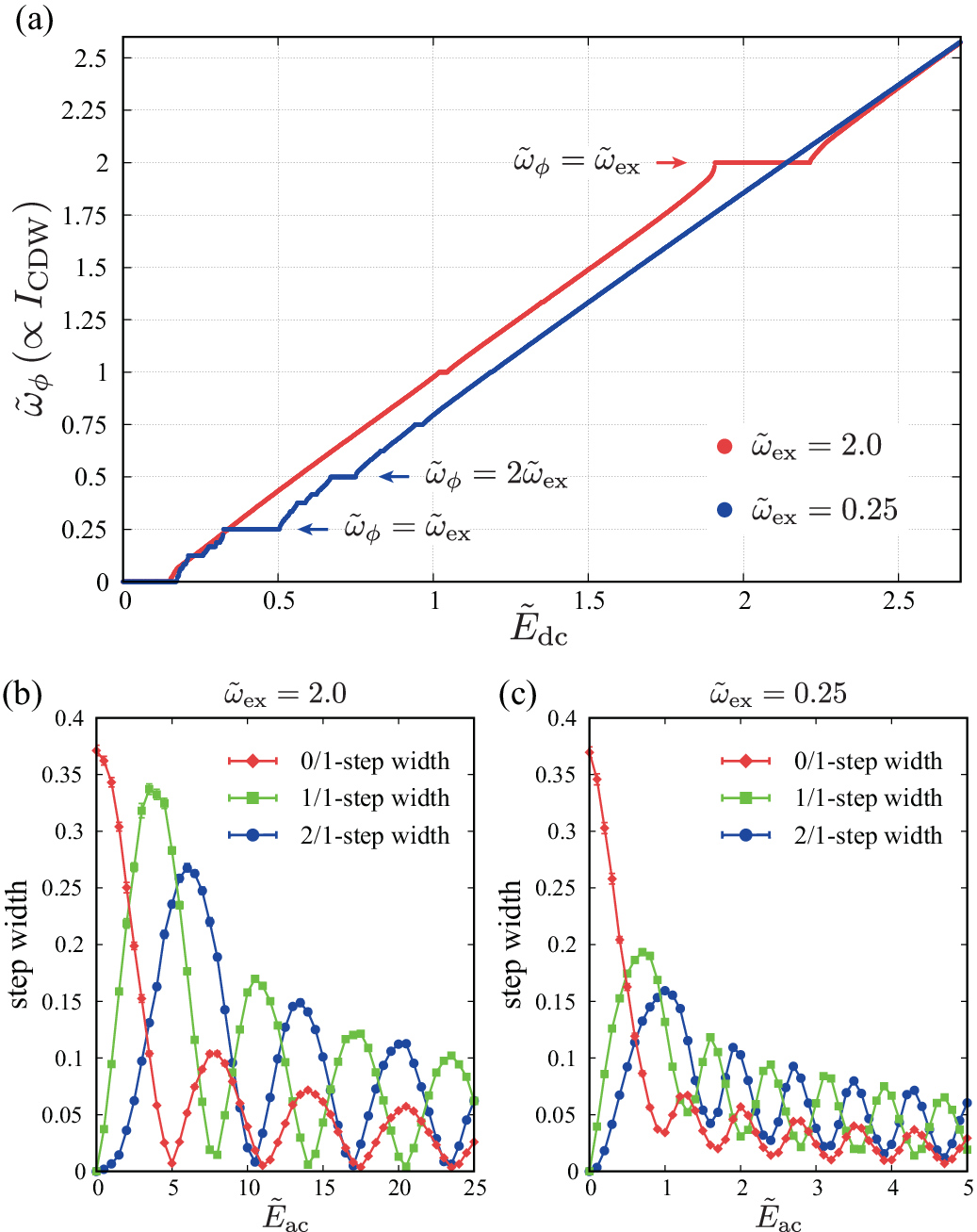}
\caption{(a) The $\tilde{E}_{\rm dc}$ dependence of $\tilde{\omega}_{\phi}$ (the $I$-$V$ characteristics) in the one-frequency case of $\tilde{E}_{\rm ac}^{\prime}=0$, where red and blue symbols represent the results obtained for $\tilde{E}_{\rm ac}=3.0$ and $\tilde{\omega}_{\rm ex}=2.0$, and $\tilde{E}_{\rm ac}=0.5$ and $\tilde{\omega}_{\rm ex}=0.25$, respectively.
The results are obtained for $\tilde{P}_{\rm imp}=1.0$ and a $\beta_i$ configuration.
(b) and (c) The $\tilde{E}_{\rm ac}$ dependence of the $0/1$-step width $W_{0/1}$ (red), $1/1$-step width $W_{1/1}$ (green), and $2/1$-step width $W_{2/1}$ (blue) for (b) $\tilde{\omega}_{\rm ex}=2.0$ and (c) $\tilde{\omega}_{\rm ex}=0.25$. 
Note that the $0/1$ step corresponds to the threshold field.
}
\label{fig:IV-single}
\end{center}
\end{figure}

Under the dc+ac fields with a one frequency $E=E_{\rm dc}+E_{\rm ac}\sin(\omega_{\rm ex}t)$, it is known that Eq. (\ref{eq:FLR-model}) leads to the occurrence of harmonic ($p/q=$ integer value) and subharmonic ($p/q=$ fractional value) Shapiro steps as observed in experiments.\cite{Richard1982, Brown1984,Thorne1987,Thorne1988,Bhattacharya1987,Higgins1993} A typical $I$-$V$ characteristic obtained from Eq. (\ref{eq:FLR-model}) is shown in Fig \ref{fig:IV-single}(a), where the plateau regions correspond to the Shapiro steps (for details, see later). Since in the overdamped regime, the occurrence of the subharmonics is due to many-body effects \cite{Gruner1988,Coppersmith1986,Matsukawa_JJAP_1987,Middleton1992} and the single-impurity model can only describe the harmonics,\cite{Gruner1981,Zettl1984} the multi-impurity FLR model used here should provide a reasonable modeling to discuss the experimental result.
In the presence of two frequencies, we simply use the following expression for the external fields:
\begin{equation}\label{two-fre}
E(t)=E_{\mathrm{dc}}+E_{\mathrm{ac}} \sin \left(\omega_{\mathrm{ex}} {t}\right)+{E}_{\mathrm{ac}}^{\prime} \sin \left(\omega_{\mathrm{ex}}^{\prime} {t}\right).
\end{equation}

\begin{figure}[t]
\begin{center}
\includegraphics[width=0.975\columnwidth]{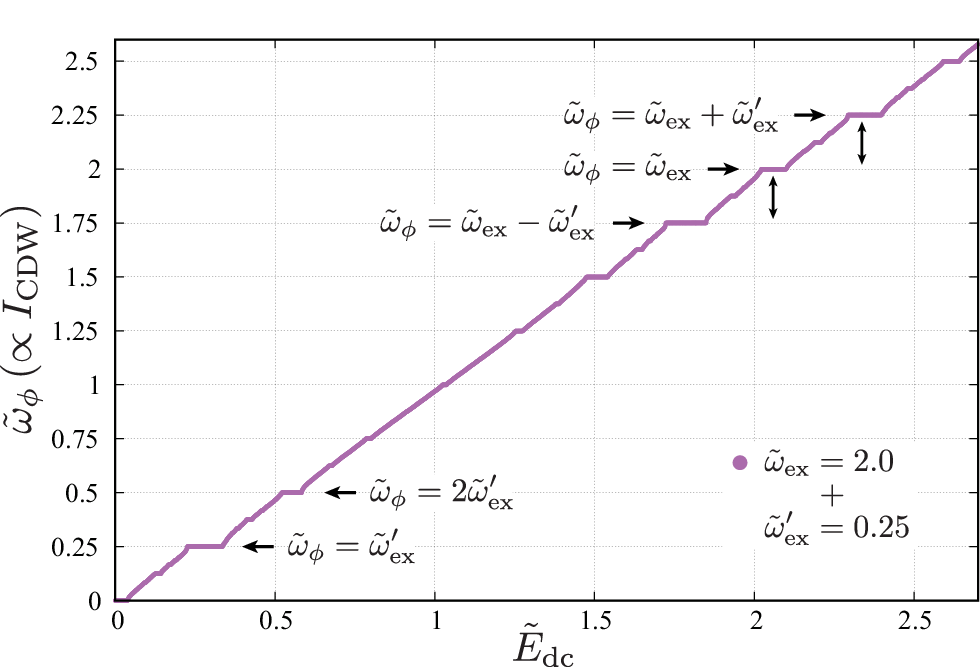}
\caption{The $\tilde{E}_{\rm dc}$ dependence of $\tilde{\omega}_{\phi}$ (the $I$-$V$ characteristics) in the two-frequency case with $\tilde{E}_{\rm ac}=3.0$, $\tilde{\omega}_{\rm ex}=2.0$, $\tilde{E}^{\prime}_{\rm ac}=0.5$, and $\tilde{\omega}^{\prime}_{\rm ex}=0.25$, where the result is obtained for $\tilde{P}_{\rm imp}=1.0$ and a fixed $\beta_i$ configuration.}
\label{fig:IV-mix}
\end{center}
\end{figure}

\begin{figure*}[t]
\begin{center}
\includegraphics[width=2\columnwidth]{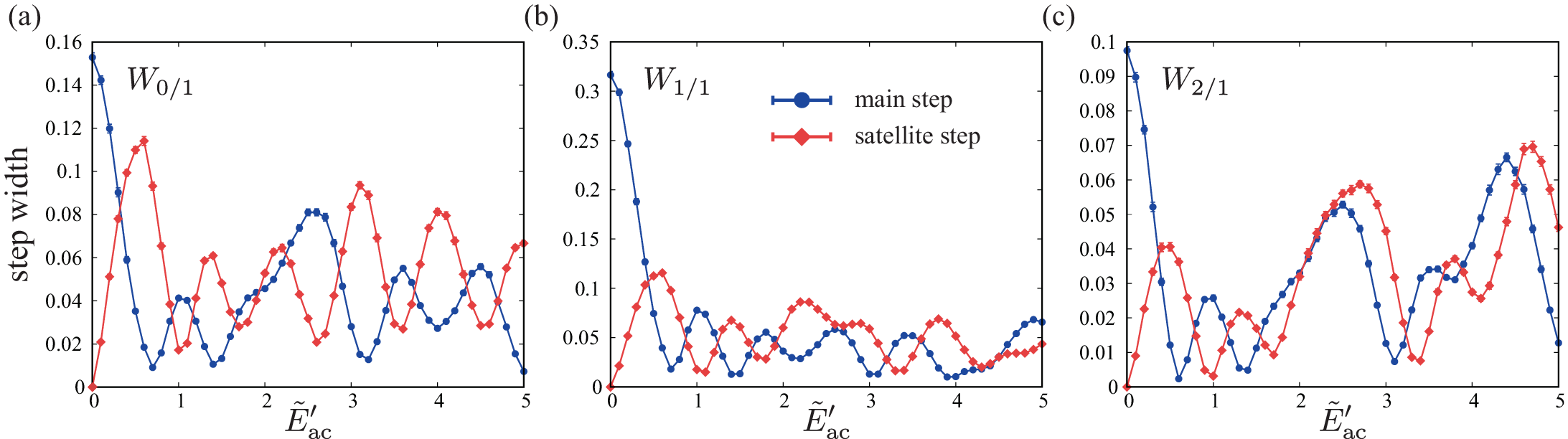}
\caption{The numerically obtained $\tilde{E}^{\prime}_{\rm ac}$ dependence of (a) the $0/1$-step width $W_{0/1}$, (b) the $1/1$-step width $W_{1/1}$, and (c) the $2/1$-step width $W_{2/1}$, where blue and red symbols correspond to the main [$\tilde\omega_\phi=(p/1)\tilde{\omega}_{\rm ex}$] and upper satellite [$\tilde{\omega}_\phi=(p/1)\tilde{\omega}_{\rm ex}+\tilde{\omega}_{\rm ex}^\prime$] steps, respectively. 
The blue symbol in (a) corresponds to the threshold field.
The same parameter set as that for Fig. \ref{fig:IV-mix}  is used except $\tilde{E}^{\prime}_{\rm ac}$.
}
\label{fig:step-simulation}
\end{center}
\end{figure*}

To obtain a steady state of $\phi(x)$ under the electric fields (\ref{two-fre}), we numerically solve Eq. (\ref{eq:FLR-model}) which can be rewritten in the dimensionless form as
\begin{equation}\label{eq:D_FLR-model}
\frac{d\phi_{i}}{d\tilde{t}}-\left(\phi_{i+1}-2\phi_{i}+\phi_{i-1}\right)=\tilde{P}_{\mathrm{imp}} \sin (\phi_i+\beta_i)+\tilde{E},
\end{equation}
where the spatial coordinate is discretized in the unit of the phase-coherence length (Fukuyama-Lee-Rice length) $L_0$,\cite{Fukuyama1978a, Tua1984, Tua1985, PRBL2023} and $\tilde{P}_{\mathrm{imp}}=P_{\mathrm{imp}}(L^2_{0}/v^{2}_{\mathrm{ph}})\sqrt{n_{\rm imp}/L_0}$ with a number density of impurities $n_{\rm imp}$,  $\tilde{E}=({eQ}L^2_{0}/{m^{\ast}} v^2_{\mathrm{ph}})E$, $\tilde{t}=t/[(\gamma\, L^2_{0})/v^2_{\mathrm{ph}}]$, and the associated frequency $\tilde{\omega}$ are dimensionless quantities.
$\beta_{i}$ represents the spatially varying optimal value of the phase and it is modeled here by a $[0, 2\pi]$ random number.
The CDW current averaged over space and time is given by  $I_{\mathrm{CDW}}\propto \frac{1}{N}\sum_{i=1}^{N} \big\langle \frac{d\phi_i}{d\tilde{t}} \big\rangle_{\tilde{t}}=\tilde{\omega}_{\phi} $ [see Eq. (\ref{CDW-current})],
where $N$ is a number of FLR domains, and $\langle \rangle_{\tilde{t}}$ denotes the time average.
We integrate Eq. (\ref{eq:D_FLR-model}) by using the fourth-order Runge-Kutta method with the time step $\Delta \tilde{t}=0.1$, typically up to $\tilde{t}=2.0\times10^5$ after discarding the first $10^4$ time steps for relaxation.
The random average over the impurity distributions corresponding to the $\beta_i$ configurations is basically taken over 30 samples.
In the two-frequency case, we use the frequency ratio $\tilde{\omega}_{\rm ex}:\tilde{\omega}^{\prime}_{\rm ex}=2.0:0.25$ ($8:1$), bearing the associated experimental ratio of 400 and 50 MHz in our mind.\cite{Nikonov2021}
For fixed values of $\tilde{E}_{\rm ac}=3.0$, $\tilde{P}_{\rm imp}= 1.0$, and $N=200$, we calculate $\tilde{\omega}_{\phi} \propto I_{\rm CDW}$ as a function of $\tilde{E}_{\rm dc}$ to obtain the $I$-$V$ characteristics. $\tilde{E}^{\prime}_{\rm ac}$ is a free parameter here.

Before discussing frequency mixing effects, we will make a brief review of the Shapiro steps in the one-frequency case of $\tilde{E}^{\prime}_{\rm ac}=0$.
Figure \ref{fig:IV-single}(a) shows a typical example of the $\tilde{E}_{\rm dc}$ dependence of $\tilde{\omega}_{\phi}$, i.e., the $I$-$V$ characteristics, for $\tilde{\omega}_{\rm ex}=2.0$ (red symbols) and $\tilde{\omega}_{\rm ex}=0.25$ (blue symbols).
In both cases, one can see the plateau regions where the CDW frequency $\tilde{\omega}_\phi$ is synchronized to $p/q$ multiple of the external frequency $\tilde{\omega}_{\rm ex}$, being robust against the driving force $\tilde{E}_{\rm dc}$. These plateaus correspond to the Shapiro steps, each characterized by $p/q$. For example, in the case of  $\tilde{\omega}_{\rm ex}=2.0 \, (0.25)$ in Fig. \ref{fig:IV-single}(a), the plateau at $\tilde{\omega}_{\phi}=2.0 \, (0.25)$ corresponds to the $p/q=1/1$ step, and in the same manner, we can label each step with $p/q$. We note that the $0/1$ step corresponds to the depinning field (threshold field).
Figure \ref{fig:IV-single}(b) and (c) show the $p/q$-step width $W_{p/q}$ as a function of the amplitude of the ac field $\tilde{E}_{\rm ac}$ for $\tilde\omega_{\rm ex}=2.0$ and $0.25$, respectively. One can see that with increasing $\tilde{E}_{\rm ac}$, the step width $W_{p/q}$ tends to be suppressed, exhibiting an oscillatory behavior. Such a damped oscillation of $W_{p/q}$ is known to be of Bessel function type with its order determined by the indices $p$ and $q$.\cite{Thorne1987}

\begin{figure*}[t]
\begin{center}
\includegraphics[width=2\columnwidth]{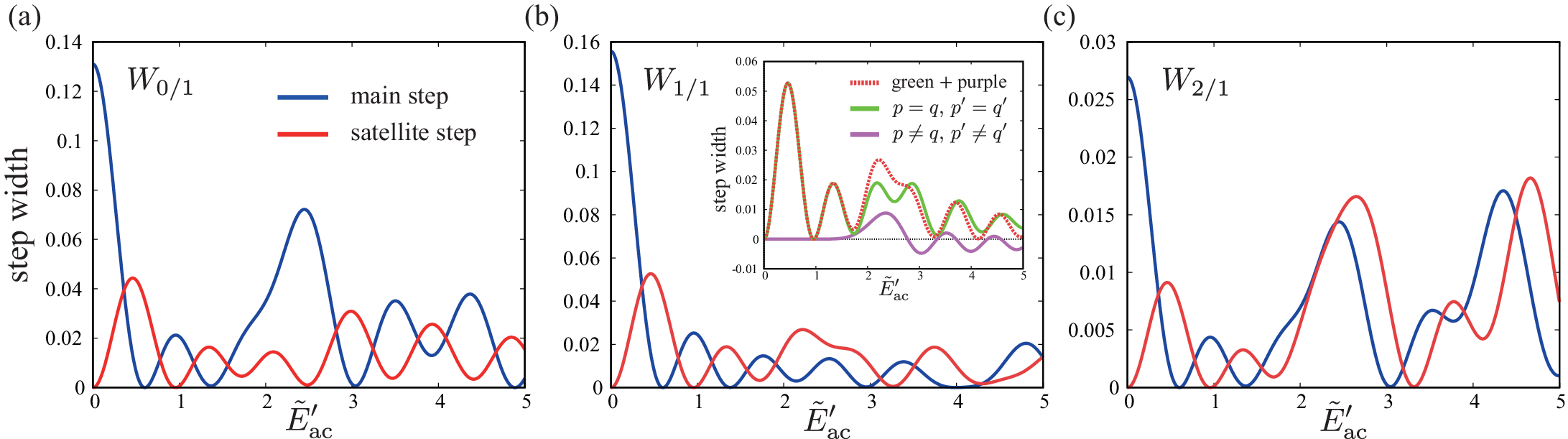}
\caption{The analytically obtained $\tilde{E}^{\prime}_{\rm ac}$ dependence of (a) the $0/1$-step width $W_{0/1}$, (b) the $1/1$-step width $W_{1/1}$, and (c) the $2/1$-step width $W_{2/1}$, where the parameters and the color notations are the same as those in Fig. \ref{fig:step-simulation}.
In (b), the inset shows the satellite-step width calculated for $|p|$, $|q|$, $|p^\prime|$, and $|q^\prime| <10$ in the evaluation of Eq. (\ref{eq:width}), where green and purple curves denote the components for $p = q$, $p^{\prime} = q^{\prime}$ and $p \neq q$, $p^{\prime} \neq q^{\prime}$, respectively, and a red dashed curve denotes their sum (see the text).
}
\label{fig:step-perturbation}
\end{center}
\end{figure*}

Now, we will turn to the two-frequency case of our interest.
Figure \ref{fig:IV-mix} shows the $I$-$V$ characteristics in the ac fields with the two frequencies $\tilde{\omega}_{\rm ex}=2.0$ and $\tilde{\omega}^{\prime}_{\rm ex}=0.25$. Comparing the one-frequency case of Fig. \ref{fig:IV-single}(a) and the two-frequency case of Fig. \ref{fig:IV-mix}, one can see that additional Shapiro steps emerge, for example, at $\tilde\omega_\phi=1.75$ and $2.25$, both branched from the original $1/1$ step of $\tilde{\omega}_\phi = (1/1)\tilde{\omega}_{\rm ex}=2.0$. Since the lower external frequency is $\tilde{\omega}^{\prime}_{\rm ex}=0.25$, these satellite steps symmetric with respect to $\tilde{\omega}_\phi = \tilde{\omega}_{\rm ex}$ can be identified with $\tilde\omega_{\phi}=\tilde\omega_{\rm ex}\pm\tilde\omega^{\prime}_{\rm ex}$. In Fig. \ref{fig:IV-mix}, higher-order satellite steps can also be found at $\tilde\omega_{\phi}=\tilde\omega_{\rm ex}\pm 2\tilde\omega^{\prime}_{\rm ex}$, i.e., $\tilde{\omega}_\phi = 1.5$ and $2.5$.
When we focus on the width of the main step at $\tilde\omega_{\phi}=2.0$, we notice that it is significantly suppressed compared with that in the one-frequency case shown in Fig. \ref{fig:IV-single}(a). To look into the details of the frequency-mixing effect on the step width, we examine the $\tilde{E}^{\prime}_{\rm ac}$ dependence of the step width.

Figure \ref{fig:step-simulation} shows the $p/1$-step widths of the main ($\tilde\omega_{\phi}=p\tilde\omega_{\rm ex}$) and its upper satellite ($\tilde\omega_{\phi}=p\tilde{\omega}_{\rm ex}+\tilde\omega_{\rm ex}^{\prime}$) steps as a function of the amplitude of the lower-frequency ac field $\tilde{E}^{\prime}_{\rm ac}$, 
where blue and red symbols represent the widths of the main and satellite steps, respectively. For comparison with the experimental data,\cite{Nikonov2021} results for the three different indices of steps are shown. We can see from Fig. \ref{fig:step-simulation} that in the small $\tilde{E}_{\rm ac}^{\prime}$ region, the step width exhibits the damped oscillation as in the one-frequency case shown in Figs. \ref{fig:IV-single}(b) and \ref{fig:IV-single}(c).
By contrast, in the large $\tilde{E}^{\prime}_{\rm ac}$ region, although the step width exhibits an oscillation, it is not monotonically damped but can be rather enhanced in all three cases shown in Fig. \ref{fig:step-simulation}, suggesting that the step formation can be promoted in the large $\tilde{E}^{\prime}_{\rm ac}$ region. Such a situation is also the case for the lower satellite (see the supplementary material).
Although the suppression of the main step in the small $\tilde{E}_{\rm ac}^{\prime}$ region can be understood as a result of the fact that the CDW mode $\tilde{\omega}_{\phi}$ synchronized to $\tilde\omega_{\rm ex}$ is disturbed by the additional oscillation $\tilde\omega_{\rm ex}^{\prime}$, the enhancement of the step width in the large $\tilde{E}_{\rm ac}^{\prime}$ region is quite non-trivial.

To understand why the satellite steps appear and the amplitude dependence of the step width does not show the conventional damped oscillation, we perform an analytical calculation where a perturbative expansion with respect to the pinning force is used.\cite{Matsukawa_Takayama, PRBL2023}
In the perturbative analysis, the Shapiro-step formation is determined by the self-consistent equation for $\tilde{\omega}_\phi$, $\tilde\omega_\phi=\tilde{E}_{\rm P}(\tilde{\omega}_\phi)+\tilde{E}_{\rm dc}$.
In the one-frequency case of $\tilde{E}_{\rm ac}^{\prime}=0$, the leading-order contribution to $\tilde{E}_{\rm P}(\tilde{\omega}_\phi)$ becomes\cite{PRBL2023}
\begin{equation}\label{eq:single-leading-order}
\tilde{E}_{\mathrm{P}}^{(2)}\left(\tilde\omega_\phi\right)\propto \frac{\tilde{P}_{\rm imp}^2}{2}\sum_p J_p^2\left(\frac{\tilde{E}_{\mathrm{ac}}}{\tilde\omega_{\mathrm{ex}}}\right) \mathrm{Im}\left[H\left(\tilde\omega_\phi-p \tilde\omega_{\mathrm{ex}}\right)\right], 
\end{equation}
where $J_p(x)$ is the Bessel function of the first kind with integer $p$. Since $H(x)$ is a function that diverges at $x=0$, the solution of the self-consistent equation is $\tilde\omega_\phi=p\tilde\omega_{\rm ex}$, and the coefficient $J_p^2\left({\tilde{E}_{\mathrm{ac}}}/{\tilde\omega_{\mathrm{ex}}}\right)$ can be interpreted as the width of the $p/1$ step.\cite{PRBL2023}
Thus, the step width exhibits a Bessel-type oscillation as a function of $({\tilde{E}_{\mathrm{ac}}}/{\tilde\omega_{\rm ex}})$.
We note in passing that higher-order contributions are relevant to the subharmonic steps.

Now we extend the perturbative approach to the two-frequency case (for details of the following calculation, see the supplementary material) where the second-order term is modified to
\begin{equation}\label{eq:two-leading-order}
\begin{aligned}
 &\tilde{E}_{\mathrm{P}}^{(2)}\left(\tilde\omega_\phi\right)\propto  \frac{\tilde{P}_{\rm imp}^2}{4}
\sum_{p, p^{\prime}, q, q^{\prime}} 
C_{pp^{\prime}qq^{\prime}}
J_p\left(\frac{\tilde{E}_{\mathrm{ac}}}{\tilde\omega_{\mathrm{ex}}}\right) J_{q}\left(\frac{\tilde{E}_{\mathrm{ac}}}{\tilde\omega_{\mathrm{ex}}}\right) \\
&\times J_{p^{\prime}}\left(\frac{\tilde{E}_{\mathrm{ac}}^{\prime}}{\tilde\omega_{\mathrm{ex}}^{\prime}}\right) J_{q^{\prime}}\left(\frac{\tilde{E}_{\mathrm{ac}}^{\prime}}{\tilde\omega_{\mathrm{ex}}^{\prime}}\right)\mathrm{Im}\left[H\left(\tilde\omega_\phi-p \tilde\omega_{\mathrm{ex}}-p^{\prime} \tilde\omega_{\mathrm{ex}}^{\prime}\right)\right],\\
&C_{pp^{\prime}qq^{\prime}} \equiv (i)^{p+p^{\prime}+q+q^{\prime}}\left\{(-1)^{p+p^{\prime}}+(-1)^{q+q^{\prime}}\right\}
\end{aligned}
\end{equation}
with integers $p$, $q$, $p^{\prime}$, and $q^{\prime}$ satisfying
\begin{equation}\label{eq:two-leading-order_delta}
(p-q) \tilde\omega_{\mathrm{ex}}+\left(p^{\prime}-q^{\prime}\right) \tilde\omega_{\mathrm{ex}}^{\prime}=0.
\end{equation}
One can see from Eq. (\ref{eq:two-leading-order}) that the synchronization condition, i.e., the condition for which the argument of $\tilde{H}(x)$ becomes zero, is a linear combination of two frequencies $\tilde\omega_\phi=p\tilde\omega_{\rm ex}+p^{\prime}\tilde\omega^{\prime}_{\rm ex}$, and this condition yields the appearance of the main and satellite steps; for example, $(p,p^\prime)=(1,0)$ and $(p,p^{\prime})=(1,\pm1)$ correspond to the $1/1$ main and satellite steps, respectively. In contrast to the one-frequency case of Eq. (\ref{eq:single-leading-order}) where the coefficient is a double product of Bessel functions, the associated coefficient in Eq. (\ref{eq:two-leading-order}) is four products of Bessel functions whose orders $p$, $q$, $p^{\prime}$, and $q^{\prime}$ satisfy the following conditions:
\begin{equation}\label{eq:condition_order}
\left\{\begin{array}{l}
\tilde\omega_\phi=p\tilde\omega_{\rm ex}+p^{\prime}\tilde\omega_{\rm ex}^{\prime}\\
\tilde\omega_\phi=q\tilde\omega_{\rm ex}+q^{\prime}\tilde\omega_{\rm ex}^{\prime}
\end{array}\right.
.
\end{equation}
The latter condition arises from Eq. (\ref{eq:two-leading-order_delta}). Thus, the quantity 
\begin{equation}\label{eq:width}
\begin{aligned}
& \overline{W}_{\tilde{\omega}_\phi}\equiv\frac{\tilde{P}_{\rm imp}^2}{4}\sum_{p, p^{\prime}, q, q^{\prime}} \delta(p, p^{\prime}, q, q^{\prime};\tilde{\omega}_\phi) C_{pp^{\prime}qq^{\prime}}\\
& \times J_p\left(\frac{\tilde{E}_{\mathrm{ac}}}{\tilde\omega_{\mathrm{ex}}}\right) J_{q}\left(\frac{\tilde{E}_{\mathrm{ac}}}{\tilde\omega_{\mathrm{ex}}}\right) J_{p^{\prime}}\left(\frac{\tilde{E}_{\mathrm{ac}}^{\prime}}{\tilde\omega_{\mathrm{ex}}^{\prime}}\right) J_{q^{\prime}}\left(\frac{\tilde{E}_{\mathrm{ac}}^{\prime}}{\tilde\omega_{\mathrm{ex}}^{\prime}}\right)
\end{aligned}
\end{equation}
corresponds to the step width for the mode $\tilde{\omega}_\phi$, where $\delta(p, p^{\prime}, q, q^{\prime};\tilde{\omega}_\phi)=1$ and $0$ for the combinations of $p$, $p^{\prime}$, $q$, and $q^{\prime}$ satisfying Eq. (\ref{eq:condition_order}) and for others, respectively.
As one can see from Eq. (\ref{eq:condition_order}), the satellite step with $p^\prime = \pm 1$ should always appears regardless of specific values of the two frequencies, although its width determined by Eq. (\ref{eq:width}) generally depends on the parameter set, as will be explained later.
Since the amplitude of the Bessel function $|J_{p}(x)|$ decreases with increasing $|p|$, the dominant contribution in Eq. (\ref{eq:width}) comes from sets of smaller $|p|$, $|q|$, $|p^{\prime}|$, and $|q^{\prime}|$. 
Figure \ref{fig:step-perturbation} shows the $\tilde{E}_{\rm ac}^{\prime}$ dependence of $\overline{W}_{p\tilde{\omega}_{\rm ex}}$ (blue symbols) and $\overline{W}_{p\tilde{\omega}_{\rm ex}+\tilde{\omega}_{\rm ex}^\prime}$ (red symbols) for (a) $p=0$, (b) $p=1$, and (c) $p=2$, where the same parameters as those for Fig. \ref{fig:step-simulation} are used. 
In the evaluation of $\overline{W}_{\tilde{\omega}_{\phi}}$,  the summation $\sum_{p,p^\prime,q,q^\prime}$ runs over $|p|,|q|,|p^{\prime}|,|q^{\prime}| < 50$ such that it definitely converges.
Since $\overline{W}_{p\tilde{\omega}_{\rm ex}}$ and $\overline{W}_{p\tilde{\omega}_{\rm ex}+\tilde{\omega}_{\rm ex}^\prime}$ correspond to the widths of the $p/1$ main and upper satellite steps, respectively, the same notations as those in Fig. \ref{fig:step-simulation} have been used. One can see from Figs. \ref{fig:step-simulation} and \ref{fig:step-perturbation}, the analytical result is qualitatively consistent with the numerical result, suggesting that in the two-frequency case, the Shapiro-step formation can be described by Eqs. (\ref{eq:condition_order}) and (\ref{eq:width}). 
As the validity of Eq. (\ref{eq:width}) is confirmed, we proceed to the origin of the step-width enhancement in the large $\tilde{E}^{\prime}_{\mathrm{ac}}$ region.
 
Since $J_p(x)$ oscillates, taking positive and negative values, their products $J_p(x)J_q(x)J_{p^\prime}(x^\prime)J_{q^\prime}(x^\prime)$ can become small or large depending on the combinations of $p$, $q$, $p^\prime$, and $q^\prime$. Here, we categorize the combinations into two; one is the combination of $p=q$ and $p^{\prime}=q^{\prime}$ for which the product $J_p^2(x)J_{p^\prime}^2(x')$ definitely takes a positive value, and the other is other combinations for which the product can take positive and negative values. As an example, we consider the satellite step branched from the $1/1$ step, i.e., $\tilde{\omega}_\phi=\tilde{\omega}_{\rm ex} + \tilde{\omega}_{\rm ex}^\prime$, picking up dominant  $J_p(x)$ components having relatively smaller indices of $|p|<10$. Noting that $\tilde{\omega}_{\rm ex}:\tilde{\omega}_{\rm ex}^\prime=8:1$, the satellite mode $1 \, \tilde{\omega}_{\rm ex} + 1 \, \tilde{\omega}_{\rm ex}^\prime$ can be expressed in different ways as $0 \, \tilde{\omega}_{\rm ex} + 9 \, \tilde{\omega}_{\rm ex}^\prime$ and $2 \, \tilde{\omega}_{\rm ex} - 7 \, \tilde{\omega}_{\rm ex}^\prime$, so that the former-category combination includes $(p,q,p^{\prime},q^{\prime})=(1,1,1,1)$, $(0,0,9,9)$, and $(2,2,-7,-7)$, whereas the latter includes $(p,q,p^{\prime},q^{\prime})=(0,2,9,-7)$ and $(2,0,-7,9)$. The inset of Fig. \ref{fig:step-perturbation}(b) shows the step width of the $1/1$ satellite for the former (green curve) and latter (purple curve) categories. Since their sum, denoted by a red dotted curve in the inset, is in good agreement with the result of the full calculation denoted by a red curve in the main panel, the contributions from the smaller $|p|$, $|q|$, $|p^\prime|$, and $|q^\prime|$ turn out to be dominant as expected. In the inset, the weak undamped behavior caused by the superposition of the synchronizations satisfying the symmetric condition $p=q$ and $p^\prime=q^\prime$ (see the green curve) is further amplified by the additional contributions from the non-symmetric one $p \neq q$ and $p^\prime \neq q^\prime$ (see the purple curve) which can take negative values, eventually leading to the enhancement of the step width in the large $\tilde{E}^\prime_{\rm ac}$ region. Thus, the complex non-monotonic $\tilde{E}^\prime_{\rm ac}$ dependence of the step width is due to the emergent multi-channel synchronization involving both frequencies.

To summarize, we have theoretically investigated the CDW sliding under the additional ac fields with two frequencies $\omega_{\rm ex}$ and $\omega_{\rm ex}^\prime$.
It is found that the CDW mode $\omega_\phi$ is synchronized to $ p \omega_{\rm ex}+ p^\prime \omega_{\rm ex}^{\prime}$ with integers $p$ and $p^\prime$, and as a result, main ($\omega_\phi= p \omega_{\rm ex}$) and satellite ($\omega_\phi= p \omega_{\rm ex}\pm\omega_{\rm ex}^\prime$) Shapiro steps appear in the $I$-$V$ characteristics. With increasing the amplitude of the ac field $E_{\rm ac}^\prime$ for the lower frequency $\omega_{\rm ex}^\prime$, the step width first exhibits a damped oscillation as in the one-frequency case, and then, exhibits a non-monotonic behavior. Such a complicated $E_{\rm ac}^\prime$ dependence is due to the multi-frequency synchronization effect, i.e., the existence of various possible combinations of $p$ and $p^\prime$ satisfying $\omega_\phi^\ast= p \omega_{\rm ex}+ p^\prime \omega_{\rm ex}^{\prime}$ for a fixed CDW mode $\omega_\phi^\ast$. 

In the associated two-frequency experiment,\cite{Nikonov2021} the main-satellite structure in the $I$-$V$ characteristics and the damped oscillatory behavior of the step width as a function of $E_{\rm ac}^\prime$ have been observed. Our theoretical results are consistent with these observations. On the other hand, in the large $E_{\rm ac}^\prime$ region where we predict the complicated non-monotonic $E_{\rm ac}^\prime$ dependence of the step width, there are no experimental data reported so far. Whether such a significant difference between the one-frequency and two-frequency cases can actually be observed or not would be an interesting question. 
Also, the concept of the frequency mixing could be extended to more than two frequencies and also be applied to other periodic driving forces such as time-dependent strain and surface acoustic waves.\cite{Nikitin2021, LT29, Mori2023, PRBL2023, SCES2023, Nikitin2023}
The extension of our theory to these platforms will be discussed elsewhere. Although associated experiments might be challenging, we believe that these issues will be investigated in future experimental studies.

See the supplementary material for the amplitude dependence of the lower-satellite-step width and details of the analytical calculation.

This work was supported by Grant-in-Aid for JSPS Fellows No. JP24KJ1638 and JSPS KAKENHI Grant Nos. JP21K03469 and JP23H00257.
One of the authors (Y.F.) was supported by Program for Leading Graduate
Schools: ``Interactive Materials Science Cadet Program''.

\newpage
\,

\pagebreak
\widetext

\begin{center}
\large\textbf{Supplementary Material for Effects of frequency mixing on Shapiro-step formations in sliding charge-density-waves}
\end{center}

\makeatletter
\setcounter{equation}{0}
\setcounter{figure}{0}
\renewcommand{\theequation}{S\arabic{equation}}
\renewcommand{\thefigure}{S\arabic{figure}}
\renewcommand{\bibnumfmt}[1]{[S#1]}
\renewcommand{\citenumfont}[1]{S#1}





\date{\today}


\maketitle 

\section{Amplitude dependence of the lower-satellite-step width}
\begin{figure}[h]
\begin{center}
\includegraphics[width=0.8\columnwidth]{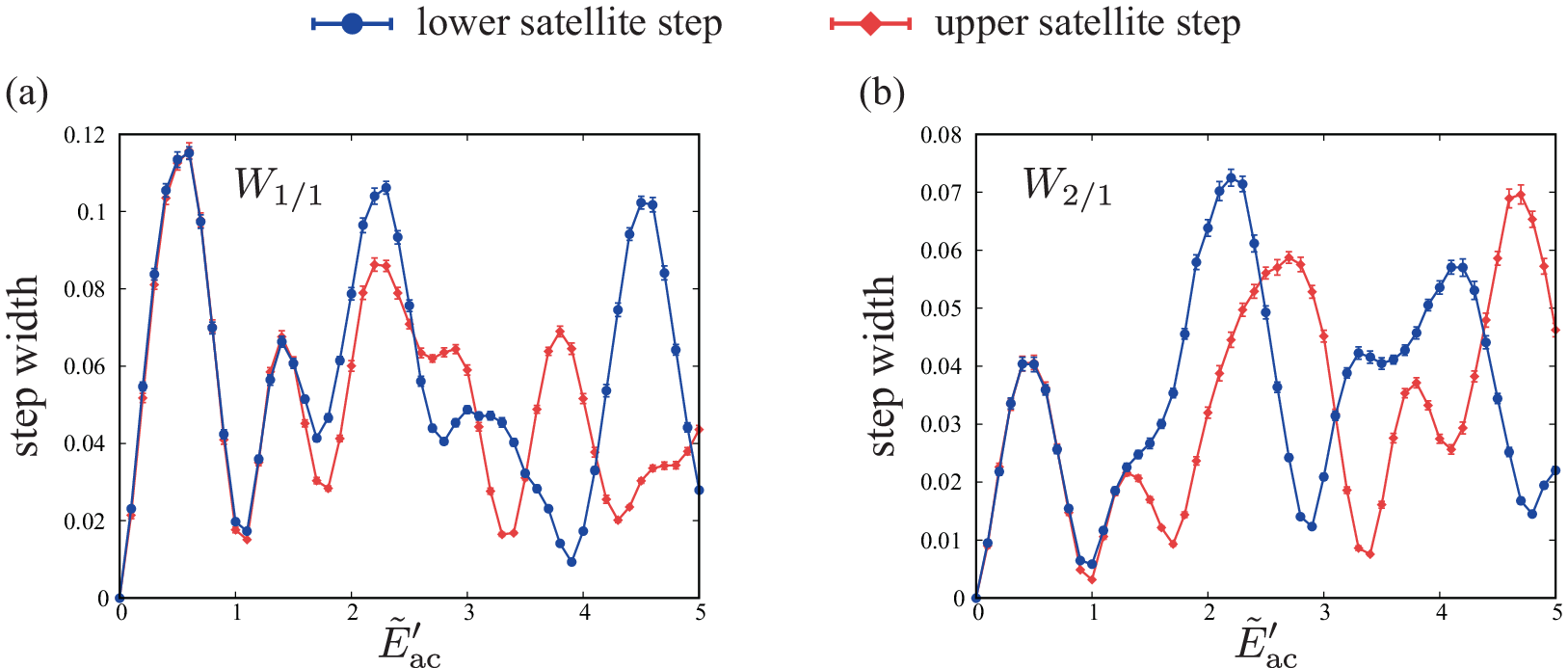}
\caption{
The numerically obtained $\tilde{E}_{\rm ac}^{\prime}$ dependence of (a) the 1/1 satellite-step width $W_{1/1}$ and (b) the 2/1 satellite-step width $W_{2/1}$,
where blue and red symbols correspond to the lower satellite [$\tilde{\omega}_\phi=(p/1)\tilde{\omega}_{\rm ex}-\tilde{\omega}_{\rm ex}^\prime$] and upper satellite [$\tilde{\omega}_\phi=(p/1)\tilde{\omega}_{\rm ex}+\tilde{\omega}_{\rm ex}^\prime$] steps, respectively. The red-symbol data are taken from those in Fig. \ref{fig:step-simulation} in the main text. The same parameter set as that for Fig. \ref{fig:step-simulation} in the main text is used.
}
\label{figS:lower_satellite}
\end{center}
\end{figure}

In the main text, we discuss the amplitude $\tilde{E}_{\rm ac}^{\prime}$ dependence of the step widths of the main [$\tilde{\omega}_\phi=(p/1)\tilde{\omega}_{\rm ex}$] and upper-satellite [$\tilde{\omega}_\phi=(p/1)\tilde{\omega}_{\rm ex}+\tilde{\omega}_{\rm ex}^\prime$] steps (see Fig. \ref{fig:step-simulation} in the main text).
In this supplementary material, for completeness, we show the associated results for lower-satellite [$\tilde{\omega}_\phi=(p/1)\tilde{\omega}_{\rm ex}-\tilde{\omega}_{\rm ex}^\prime$] steps.
Figure \ref{figS:lower_satellite} shows the $p/1$ satellite-step widths as a function of the amplitude $\tilde{E}_{\rm ac}^{\prime}$ for (a) $p=1$ and (b) $p=2$, where blue and red symbols represent the results for the lower and upper satellites, respectively. Note that the red-symbol data are the same as those in Fig. \ref{fig:step-simulation}(b) and (c) in the main text.
As one can see from Fig. \ref{figS:lower_satellite}, both the upper and lower satellite steps do not show simple damped oscillations.

\section{Perturbative analysis in the two-frequency case}

In the main text, we discuss the Shapiro-step width based on an analytical calculation in which a perturbation with respect to the pinning force is used. Although the perturbative approach has already been discussed in the two-frequency case as well as the one-frequency case in Ref. \cite{SMatsukawa_Takayama}, here, we provide, for completeness, the details of the perturbative calculation in the two-frequency case of $E=E_{\mathrm{dc}}+E_{\mathrm{ac}} \sin \left(\omega_{\mathrm{ex}} {t}\right)+{E}_{\mathrm{ac}}^{\prime} \sin \left(\omega_{\mathrm{ex}}^{\prime} {t}\right)$.

We first derive the self-consistent equation for $\omega_\phi$, $\tilde\omega_\phi = \tilde{E}_{\rm P}(\tilde{\omega}_\phi)+\tilde{E}_{\rm dc}$, used in the main text.
In Eq. (\ref{eq:FLR-model}) in the main text, we introduce $\bar\gamma\equiv\gamma/v^2_{\rm ph}$, $\bar{E}_{\rm P}\equiv\left(P_{\rm imp}/v^2_{\rm ph}\right)N_{\rm p}(x)\sin(\phi+Qx)$, and $\bar{E}\equiv(eQ/m^{\ast}\gamma)E$ for later convenience, and rewrite Eq. (\ref{eq:FLR-model}) in the main text as follows:
\begin{equation}\label{EqS:EOM}
	\left(\bar{\gamma}\frac{\partial}{\partial t} - \nabla^2\right)\phi(x,t)=\bar{E}_{\rm P}(x,t)+\bar\gamma\bar{E}(t).
\end{equation}
In the sliding region, we could express the CDW phase $\phi(x,t)$ as $\phi(x,t)=\phi_0(t)+\delta\phi(x,t)$ with the local deviation $\delta\phi$ from a global sliding mode $\phi_0$.
The global sliding mode $\phi_0(t)$ is determined from Eq. (\ref{EqS:EOM}) without the pinning term as
\begin{equation}\label{EqS:initial_phi}
	\phi_0(t)=\omega_\phi t - \frac{\bar{E}_{\rm ac}}{\omega_{\rm ex}}\cos (\omega_{\rm ex} t) - \frac{\bar{E}_{\rm ac}^{\prime}}{\omega_{\rm ex}^{\prime}}\cos (\omega_{\rm ex}^{\prime} t).
\end{equation}
In the case of $\bar{E}_{\rm ac}^{\prime}=0$, Eq. (\ref{EqS:initial_phi}) is reduced to $\phi_0$ in the one-frequency case.\cite{SPRBL2023}
By averaging Eq. (\ref{EqS:EOM}) over space and time, we have
\begin{equation}\label{EqS:Self-consistent}
	\omega_{\phi} = \frac{1}{\bar{\gamma}}\left\langle\bar{E}_{\rm P}(x,t)\right\rangle_{x,t}+\bar{E}_{\rm dc},
\end{equation}
where $\langle A \rangle_{x,t} = \frac{1}{LT}\int dx dt$ represents the spatial-time average of $A$ and $\langle\dot{\delta\phi}(x,t)\rangle_{x,t}=0$ has been assumed.
Since $\langle\bar{E}_{\rm P}(x,t)\rangle_{x,t}$ depends on the CDW phase $\phi$, Eq. (\ref{EqS:Self-consistent}) is the self-consistent equation for $\omega_\phi$.

By substituting $\phi(x,t)=\phi_0(t)+\delta\phi(x,t)$ and Eq. (\ref{EqS:Self-consistent}) into Eq. (\ref{EqS:EOM}), we obtain
\begin{equation}\label{EqS:EOM_delta}
	\left(\bar{\gamma}\frac{\partial}{\partial t} - \nabla^2\right)\delta\phi(x,t)=\bar{E}_{\rm P}(x,t) - \left\langle \bar{E}_{\rm P}(x,t) \right\rangle_{x,t}.
\end{equation}
Then, $\delta\phi(x,t)$ can be determined by
\begin{equation}
\begin{aligned}
\delta \phi(x, t) =\int d x^{\prime} d t^{\prime} \tilde{G}\left(x-x^{\prime}, t-t^{\prime}\right) \bar{E}_{\mathrm{P}}\left(x^{\prime}, t^{\prime}\right),
\end{aligned}
\end{equation}
where $\tilde{G}(x,t)$ is defined by $\tilde{G}(x,t)\equiv G(x,t) - \langle G(x,t) \rangle_{x,t}$ with the Green function $G(x, t)$ satisfying
\begin{equation}
\left(\bar{\gamma} \frac{\partial}{\partial t}-\nabla^2\right) G(x, t)=\delta(x) \delta(t).
\end{equation}
The Fourier transform of $G(x,t)$ is
\begin{equation}
	G(k, \omega) = \frac{1}{i\bar\gamma\omega+k^2},
\end{equation}
and then, $\tilde{G}(k,\omega)=G(k,\omega)-G(0,0)\delta(k)\delta(\omega)$.
Next, we expand the pinning term $\bar{E}_{\rm P}$ with respect to $\delta \phi$ as follows:
\begin{equation}\label{EqS:E_expanded}
\begin{aligned}
	\bar{E}_{\rm P}(x,t)&=\frac{P_{\rm imp}}{v_{\rm ph}^2}N_{\rm p}(x)\sum_{n=0}^{\infty}\frac{1}{n!}\sin\left(\phi_0(t)+Qx+\frac{n}{2}\pi\right)\left[\delta\phi(x,t)\right]^n\\
	&\equiv\sum_{n=0}^{\infty}\bar{E}_{\rm P}^{(n+1)}(x,t),
\end{aligned}
\end{equation}
where the $\delta \phi^{(n)}$ is given by
\begin{equation}
	\delta\phi^{(n)}(x,t) = \int dx^{\prime} dt^{\prime}\tilde{G}(x-x^{\prime}, t-t^{\prime})\bar{E}_{\rm P}^{(n)}(x^{\prime}, t^{\prime}).
\end{equation}
Here, $\bar{E}_{\rm P}^{(n+1)}(x,t)$ and $\delta \phi^{(n)}(x,t)$ denote the $n$-th order contributions in $P_{\rm imp}$.
By substituting Eq. (\ref{EqS:E_expanded}) into Eq. (\ref{EqS:Self-consistent}), we have
\begin{equation}
	\omega_\phi = \frac{1}{\bar{\gamma}}\sum_{n=0}^{\infty}\left\langle\bar{E}_{\rm P}^{(n+1)}(x,t)\right\rangle_{x,t}+\bar{E}_{\rm dc}.
\end{equation}

Now, the problem is reduced to calculate the concrete expression of $\left\langle\bar{E}_{\rm P}^{(n+1)}(x,t)\right\rangle_{x,t}$.
The first- and second-order terms in Eq. (\ref{EqS:E_expanded}) after averaging over space can be expressed as
\begin{equation}
\begin{aligned}
\left\langle\bar{E}_{\mathrm{P}}^{(1)}(x, t)\right\rangle_x & =\frac{P_{\mathrm{imp}}}{v_{\mathrm{ph}}^2}\left\langle N_{\rm p}(x)\sin\left(\phi_0(t)+Qx\right)\right\rangle_x=\frac{P_{\mathrm{imp}}}{v_{\mathrm{ph}}^2 L} \sum_{i=1}^{N_{\rm imp}} \sin\left(\phi_0(t)+QR_i\right), \\
\left\langle\bar{E}_{\mathrm{P}}^{(2)}(x, t)\right\rangle_x & 
=\frac{P_{\mathrm{imp}}^2}{v_{\mathrm{ph}}^4 }\left\langle \int dx_1 d t_1 N_{\rm p}(x)N_{\rm p}(x_1) \tilde{G}\left(x-x_1, t-t_1\right) \cos\left(\phi_0(t)+Qx\right)\sin\left(\phi_0(t_1)+Qx_1\right)\right\rangle_x\\
&=\frac{P_{\mathrm{imp}}^2}{v_{\mathrm{ph}}^4 L} \sum_{i, j=1}^{N_{\rm imp}} \int d t_1 \tilde{G}\left(R_i-R_j, t-t_1\right) \cos\left(\phi_0(t)+QR_i\right)\sin\left(\phi_0(t_1)+QR_j\right) \\
& =\frac{P_{\mathrm{imp}}^2}{v_{\mathrm{ph}}^4 L} \sum_{i, j=1}^{N_{\rm imp}} \int d t_1 \int \frac{d k}{2 \pi} \tilde{G}\left(k, t-t_1\right) e^{i k\left(R_i-R_j\right)} \cos\left(\phi_0(t)+QR_i\right)\sin\left(\phi_0(t_1)+QR_j\right) ,
\end{aligned}
\end{equation}
where $N_{\rm imp}$ is the number of impurities.
Since the pinning sites $R_i$ are randomly distributed, $\frac{1}{L} \sum_{i=1}^{N_{\rm imp}} f(R_i)$ can be considered the random average of the function $f(x)$.
Therefore, the random average of the first-order term $\sin(\phi_0(t) +QR_i)$ vanishes, i.e., $\left\langle\bar{E}_{\mathrm{P}}^{(1)}(x, t)\right\rangle_x = 0$. The second-order term $\left\langle\bar{E}_{\mathrm{P}}^{(2)}(x, t)\right\rangle_x$, on the other hand, becomes nonzero when $R_j = R_i$, and after the random average, it reads
\begin{equation}\label{EqS:E_P2_x}
\begin{aligned}
\left\langle\bar{E}_{\mathrm{P}}^{(2)}(x, t)\right\rangle_x &  = \frac{P_{\rm imp}^2 N_{\rm imp}}{v_{\rm ph}^4 L} \int d t_1\int \frac{d k}{2 \pi}  \tilde{G}\left(k, t-t_1\right) \frac{1}{2} \sin \left(-\phi_0(t)+\phi_0\left(t_1\right)\right) \\
& = \frac{P_{\rm imp}^2 N_{\rm imp}}{v_{\rm ph}^4 L } \int d t_1 \int \frac{d k}{2 \pi}  \int \frac{ d \omega}{2 \pi} \tilde{G}(k, \omega) e^{i \omega\left(t-t_1\right)} \frac{i}{4}\left[e^{i\left\{\phi_0(t)-\phi_0\left(t_1\right)\right\}}-e^{-i\left\{\phi_0(t)-\phi_0\left(t_1\right)\right\}}\right].
\end{aligned}
\end{equation}
Substituting Eq. (\ref{EqS:initial_phi}) into Eq. (\ref{EqS:E_P2_x}) and taking the time average, we obtain
\begin{equation} \label{EqS:space-time_average}
\begin{aligned}
\left\langle\bar{E}_{\mathrm{P}}^{(2)}(x, t)\right\rangle_{x, t} & = \frac{P_{\rm imp}^2 N_{\rm imp} }{v_{\rm ph}^4 TL}\int d t  \int d t_1  \int \frac{d k}{2 \pi} \int \frac{ d \omega}{2 \pi} \tilde{G}(k, \omega) e^{i \omega\left(t-t_1\right)} \frac{i}{4}\left[e^{i\left\{\phi_0(t)-\phi_0\left(t_1\right)\right\}}-\mathrm{c.c.}\right] \\
 &\propto P_{\rm imp}^2 \int \frac{d k}{2 \pi} \int \frac{ d \omega}{2 \pi} \tilde{G}(k, \omega) \sum_{p, p^{\prime}, q, q^{\prime}} J_p\left(\frac{\bar{E}_{\mathrm{ac}}}{\omega_{\mathrm{ex}}}\right) J_q \left(\frac{\bar{E}_{\mathrm{ac}}}{\omega_{\mathrm{ex}}}\right) J_{p^{\prime}}\left(\frac{\bar{E}_{\mathrm{ac}}^{\prime}}{\omega_{\mathrm{ex}}^{\prime}}\right) J_{q^{\prime}}\left(\frac{\bar{E}_{\mathrm{ac}}^{\prime}}{\omega_{\mathrm{ex}}^{\prime}}\right) \\
& \quad\times\int d t \int d t_1 \frac{i}{4} e^{i\omega(t-t_1)} \left[(-i)^{p+p^{\prime}} (i)^{q+q^{\prime}}e^{i\left\{\left(\omega_\phi-p \omega_{\mathrm{ex}}-p^{\prime} \omega_{\mathrm{ex}}^{\prime}\right) t-\left(\omega_\phi-q\omega_{\mathrm{ex}}-q^{\prime} \omega_{\mathrm{ex}}^{\prime}\right) t_1\right\}}
-\mathrm{c.c.}
\right]
\end{aligned}
\end{equation}
with integers $p$, $q$, $p^{\prime}$, and $q^{\prime}$, where we have used the following formula for the Bessel function of the 1st kind $J_p(x)$ 
\begin{equation}
	e^{\pm i \frac{\bar{E}_{\rm ac}}{\omega_{\rm ex}}\cos\left(\omega_{\rm ex}t\right)} = \sum_{p}\left(\pm i \right)^{p}J_{p}\left(\frac{\bar{E}_{\rm ac}}{\omega_{\rm ex}}\right)e^{\pm i p \omega_{\rm ex} t}.
\end{equation}
Then, we can perform the following integrals over $t$, $t_1$, and $\omega$,
\begin{equation}
\begin{aligned}
& \int d t_1 \int d \omega \int d t \tilde{G}(k, \omega) e^{i \omega\left(t-t_1\right)} e^{i\left\{\left(\omega_\phi-p \omega_{\mathrm{ex}}-p^{\prime} \omega_{\mathrm{ex}}^{\prime}\right) t-\left(\omega_\phi-q \omega_{\mathrm{ex}}-q^{\prime} \omega_{\mathrm{ex}}^{\prime}\right) t_1\right\}} \\
& = 2\pi  \int d t_1\int d \omega \tilde{G}(k, \omega) \delta\left(\omega+\omega_\phi-p \omega_{\mathrm{ex}}-p^{\prime} \omega_{\mathrm{ex}}^{\prime}\right) e^{-i\left(\omega+\omega_\phi-q \omega_{\mathrm{ex}}-q^{\prime}\omega_{\mathrm{ex}}^{\prime}\right) t_1} \\
& = 2\pi\int d t_1 \tilde{G}\left(k, -\omega_\phi+p \omega_{\mathrm{ex}}+p^{\prime} \omega_{\mathrm{ex}}^{\prime}\right) e^{-i\left\{(p-q) \omega_{\mathrm{ex}}+\left(p^{\prime}-q^{\prime}\right) \omega_{\mathrm{ex}}^{\prime}\right\} t_1} \\
& = \left(2\pi \right)^2 \delta\Bigl((p-q) \omega_{\mathrm{ex}}+\left(p^{\prime}-q^{\prime}\right) \omega_{\mathrm{ex}}^{\prime}\Bigr) \tilde{G}\left(k, -\omega_\phi+p \omega_{\mathrm{ex}}+p^{\prime} \omega_{\mathrm{ex}}^{\prime}\right).
\end{aligned}
\end{equation}
Finally, Eq. (\ref{EqS:space-time_average}) becomes
\begin{equation}\label{EqS:E2_P}
\begin{aligned}
\left\langle\bar{E}_{\mathrm{P}}^{(2)}(x, t)\right\rangle_{x, t}&\propto P_{\rm imp}^2  \sum_{p, p^{\prime}, q, q^{\prime}} \delta\Bigl((p-q) \omega_{\mathrm{ex}}+\left(p^{\prime}-q^{\prime}\right)  \omega_{\mathrm{ex}}^{\prime}\Bigr)J_p\left(\frac{\bar{E}_{\mathrm{ac}}}{\omega_{\mathrm{ex}}}\right) J_q \left(\frac{\bar{E}_{\mathrm{ac}}}{\omega_{\mathrm{ex}}}\right) J_{p^{\prime}}\left(\frac{\bar{E}_{\mathrm{ac}}^{\prime}}{\omega_{\mathrm{ex}}^{\prime}}\right) J_{q^{\prime}}\left(\frac{\bar{E}_{\mathrm{ac}}^{\prime}}{\omega_{\mathrm{ex}}^{\prime}}\right) \\
& \quad\times \int \frac{d k}{2 \pi} \frac{i}{4}\left[
(-i)^{p+p^{\prime}} (i)^{q+q^{\prime}}\tilde{G}\left(k, -\omega_\phi+p \omega_{\mathrm{ex}}+p^{\prime} \omega_{\mathrm{ex}}^{\prime}\right)
-
(i)^{p+p^{\prime}} (-i)^{q+q^{\prime}}\tilde{G}\left(k, \omega_\phi-p \omega_{\mathrm{ex}}-p^{\prime} \omega_{\mathrm{ex}}^{\prime}\right) 
\right]\\
&= \frac{P^2_{\mathrm{imp}}}{4}\sum_{p, p^{\prime}, q, q^{\prime}} \delta\Bigl( (p-q) \omega_{\mathrm{ex}}+\left(p^{\prime}-q^{\prime}\right)  \omega_{\mathrm{ex}}^{\prime}\Bigr)
 J_p\left(\frac{\bar{E}_{\mathrm{ac}}}{\omega_{\mathrm{ex}}}\right) J_q\left(\frac{\bar{E}_{\mathrm{ac}}}{\omega_{\mathrm{ex}}}\right) J_{p^{\prime}}\left(\frac{\bar{E}_{\mathrm{ac}}^{\prime}}{\omega_{\mathrm{ex}}^{\prime}}\right) J_{q^{\prime}}\left(\frac{\bar{E}_{\mathrm{ac}}^{\prime}}{\omega_{\mathrm{ex}}^{\prime}}\right) \\
&\quad\times (i)^{p+p^{\prime}+q+q^{\prime}}\left\{(-1)^{p+p^{\prime}}+(-1)^{q+q^{\prime}}\right\}\operatorname{Im}\left\{\tilde{H}\left(\omega_\phi-p \omega_{\mathrm{ex}}-p^{\prime} \omega_{\mathrm{ex}}^{\prime}\right)\right\}.
\end{aligned}
\end{equation}
The factor $(i)^{p+p^{\prime}+q+q^{\prime}}\left\{(-1)^{p+p^{\prime}}+(-1)^{q+q^{\prime}}\right\}$ in Eq. (\ref{EqS:E2_P}) is defined as $C_{pp^\prime q q^\prime}$ in the main text. When $p+p^{\prime}+q+q^{\prime}$ is an odd integer, $\left\langle\bar{E}_{\mathrm{P}}^{(2)}(x, t)\right\rangle_{x, t}$ vanishes as it contains $\left\{(-1)^{p+p^{\prime}}+(-1)^{q+q^{\prime}}\right\}$. 
By the same consideration, it turns out that the real part of $\tilde{H}(\omega)$ vanishes, so that in Eq. (\ref{EqS:E2_P}), $\left\langle\bar{E}_{\mathrm{P}}^{(2)}(x, t)\right\rangle_{x, t}$ involves only the imaginary part of $\tilde{H}(\omega)$.
Here, we have defined $\tilde{H}(\omega)$ as $\tilde{H}(\omega) \equiv \int \frac{d k}{2 \pi} \tilde{G}(k, \omega) \equiv \mathcal{F}^{(+)}(\omega) + i \mathcal{F}^{(-)}(\omega)$, where $\mathcal{F}^{(\pm)}(\omega)$ are given by
\begin{equation}\label{EqS:F_pm}
\mathcal{F}^{( \pm)}(\omega)=\frac{(\bar{\gamma} \omega)^{-1 / 2}}{4 \sqrt{2} \pi}\left\{f_l(\omega) \pm 2 f_t\left(\frac{\sqrt{2 \bar{\gamma} \omega} \Lambda}{\bar{\gamma} \omega-\Lambda^2}\right)\right\}
\end{equation}
with a cutoff parameter $\Lambda$, and the functions $f_l(x)$ and $f_t(x)$ are defined by
\begin{equation}
f_l(x) \equiv \log \left|\frac{\Lambda^2-\sqrt{2 \bar{\gamma} x} \Lambda+\bar{\gamma} x}{\Lambda^2+\sqrt{2 \bar{\gamma} x} \Lambda+\bar{\gamma} x}\right|, \quad f_t(x) \equiv \begin{cases}\tan ^{-1} x & x \geq 0 \\ \pi+\tan ^{-1} x & x<0 .\end{cases}
\end{equation}

As can be seen from Eq. (\ref{EqS:F_pm}), each component of $\tilde{H}(\omega)$, namely $\mathcal{F}^{(\pm)}(\omega)$, diverges at $\omega=0$, so Eq. (\ref{EqS:E2_P}) also diverges at $\omega_\phi=p \omega_{\mathrm{ex}}+p^{\prime} \omega_{\mathrm{ex}}^{\prime}$.
Therefore, the solution of the self-consistent equation (\ref{EqS:Self-consistent}) is definitely $\omega_\phi=p \omega_{\mathrm{ex}}+p^{\prime} \omega_{\mathrm{ex}}^{\prime}$, which corresponds to the synchronization condition for the Shapiro steps.


\begin{thebibliography}{80}
\bibitem{Gruner1988} G. Grüner, Rev. Mod. Phys. {\bf 60}, 1129 (1988).
\bibitem{Monceau2012} P. Monceau, Adv. Phys. {\bf 61}, 325 (2012).
\bibitem{Thorne1996} R. E. Thorne, Phys. Today {\bf 49}(5), 42 (1996).
\bibitem{Zybtsev2024} S. G. Zybtsev, V. Ya. Pokrovskii, S. A. Nikonov, M. V. Nikitin, A. A. Maizlakh, A. V. Snezhko, V. V. Pavlovskiy, and S. V. Zaitsev-Zotov, JETP Lett. {\bf 119}, 123 (2024)
\bibitem{Balandin2021} A. A. Balandin, S. V. Zaitsev-Zotov, and G. Grüner, Appl. Phys. Lett. {\bf 119}, 170401 (2021).
\bibitem{Lee1974} P. A. Lee, T. M. Rice, and P. W. Anderson, Solid State Commun. {\bf 14}, 703 (1974).
\bibitem{Liu2021} L. Liu, C. Zhu, Z. Y. Liu, H. Deng, X. B. Zhou, Y. Li, Y. Sun, X. Huang, S. Li, X. Du, Z. Wang, T. Guan, H. Mao, Y. Sui, R. Wu, J.-X. Yin, J.-G. Cheng, and S. H. Pan, Phys. Rev. Lett. {\bf 126}, 256401 (2021).
\bibitem{Frohlich1954} H. Fröhlich, Proc. Roy. Sos. A {\bf 223}, 296 (1954).
\bibitem{Monceau1976} P. Monçeau, N. P. Ong, A. M. Portis, A. Meerschaut, and J. Rouxel, Phys. Rev. Lett. {\bf 37}, 602 (1976).
\bibitem{Ong1977} N. P. Ong and P. Monceau, Phys. Rev. B {\bf 16}, 3443 (1977).
\bibitem{Fleming1979} R. M. Fleming and C. C. Grimes, Phys. Rev. Lett. {\bf 42}, 1423 (1979).
\bibitem{Thorne1987a} R. E. Thorne, W. G. Lyons, J. W. Lyding, J. R. Tucker, and J. Bardeen, Phys. Rev. B {\bf 35}, 6348 (1987).
\bibitem{Reichhardt2017} C. Reichhardt and C. J. Olson Reichhardt, Rep. Prog. Phys. {\bf 80}, 026501 (2017).
\bibitem{Matsukawa1994} H. Matsukawa and H. Fukuyama, Phys. Rev. B {\bf 49}, 17286 (1994).
\bibitem{Maeda2005} A. Maeda, Y. Inoue, H. Kitano, S. Savel’ev, S. Okayasu, I. Tsukada, and F. Nori, Phys. Rev. Lett. {\bf 94}, 077001 (2005).
\bibitem{Bhattacharya1993} S. Bhattacharya and M. J. Higgins, Phys. Rev. Lett. {\bf 70}, 2617 (1993).
\bibitem{Kaji2022} T. Kaji, S. Maegochi, K. Ienaga, S. Kaneko, and S. Okuma, Sci Rep {\bf 12}, 1542 (2022).
\bibitem{Zettl1984} A. Zettl and G. Grüner, Phys. Rev. B {\bf 29}, 755 (1984).
\bibitem{McCarten1994} J. McCarten, D. A. DiCarlo, and R. E. Thorne, Phys. Rev. B \textbf{49}, 10113 (1994). 
\bibitem{Zybtsev2020} S. G. Zybtsev, S. A. Nikonov, V. Ya. Pokrovskii, V. V. Pavlovskiy, and D. Starešinić, Phys. Rev. B {\bf 101}, 115425 (2020).
\bibitem{Thorne1987} R. E. Thorne, W. G. Lyons, J. W. Lyding, J. R. Tucker, and J. Bardeen, Phys. Rev. B {\bf 35}, 6360 (1987).
\bibitem{Thorne1988} R. E. Thorne, J. S. Hubacek, W. G. Lyons, J. W. Lyding, and J. R. Tucker, Phys. Rev. B {\bf 37}, 10055 (1988).
\bibitem{Richard1982} J. Richard, P. Monceau, and M. Renard, Phys. Rev. B {\bf 25}, 948 (1982).
\bibitem{Bhattacharya1987} S. Bhattacharya, J. P. Stokes, M. J. Higgins, and R. A. Klemm, Phys. Rev. Lett. {\bf 59}, 1849 (1987).
\bibitem{Brown1984} S. E. Brown, G. Mozurkewich, and G. Grüner, Phys. Rev. Lett. {\bf 52}, 2277 (1984).
\bibitem{Higgins1993} M. J. Higgins, A. A. Middleton, and S. Bhattacharya, Phys. Rev. Lett. {\bf 70}, 3784 (1993).
\bibitem{Coppersmith1986} S. N. Coppersmith and P. B. Littlewood, Phys. Rev. Lett. {\bf 57}, 1927 (1986).
\bibitem{Matsukawa_JJAP_1987} H. Matsukawa and H. Takayama, Jpn. J. Appl. Phys. {\bf 26}, 601 (1987). 
\bibitem{Middleton1992} A. A. Middleton, O. Biham, P. B. Littlewood, and P. Sibani, Phys. Rev. Lett. {\bf 68}, 1586 (1992).
\bibitem{Nikonov2021} S. A. Nikonov, S. G. Zybtsev, and V. Ya. Pokrovskii, Appl. Phys. Lett. {\bf 118}, 253108 (2021).
\bibitem{Thorne2005} R. E. Thorne, J. Phys. IV France {\bf 131}, 89 (2005).
\bibitem{Fukuyama1976} H. Fukuyama, J. Phys. Soc. Jpn. {\bf 41}, 513 (1976).
\bibitem{Fukuyama1978} H. Fukuyama and P. A. Lee, Phys. Rev. B {\bf 17}, 535 (1978).
\bibitem{Lee1979} P. A. Lee and T. M. Rice, Phys. Rev. B {\bf 19}, 3970 (1979).
\bibitem{Zettl1982} A. Zettl, C. M. Jackson, and G. Grüner, Phys. Rev. B {\bf 26}, 5773 (1982).
\bibitem{Sridhar1985} S. Sridhar, D. Reagor, and G. Gruner, Phys. Rev. Lett. {\bf 55}, 1196 (1985).
\bibitem{Reagor1986} D. Reagor, S. Sridhar, and G. Gruner, Phys. Rev. B {\bf 34}, 2212 (1986).
\bibitem{Sridhar1986} S. Sridhar, D. Reagor, and G. Gruner, Phys. Rev. B {\bf 34}, 2223 (1986).
\bibitem{Gruner1981} G. Grüner, A. Zawadowski, and P. M. Chaikin, Phys. Rev. Lett. {\bf 46}, 511 (1981).
\bibitem{Fukuyama1978a} H. Fukuyama, J. Phys. Soc. Jpn. {\bf 45}, 1474 (1978).
\bibitem{Tua1984} P. F. Tua and A. Zawadowski, Solid State Commun. {\bf 49}, 19 (1984).
\bibitem{Tua1985} P. F. Tua and J. Ruvalds, Solid State Commun. {\bf 54}, 471 (1985).
\bibitem{PRBL2023} Y. Funami and K. Aoyama, Phys. Rev. B {\bf 108}, L100508 (2023).
\bibitem{Matsukawa_Takayama} H. Matsukawa and H. Takayama, J. Phys. Soc. Jpn. {\bf 56}, 1507 (1987); H. Matsukawa, {\it ibid}. {\bf 56}, 1522 (1987); {\bf 57}, 3463 (1988).
\bibitem{Nikitin2021} M. V. Nikitin, S. G. Zybtsev, V. Ya. Pokrovskii, and B. A. Loginov, Appl. Phys. Lett. {\bf 118}, 223105 (2021).
\bibitem{LT29} Y. Funami and K. Aoyama, JPS Conf. Proc. {\bf 38}, 011059 (2023).
\bibitem{Mori2023} M. Mori and S. Maekawa, Appl. Phys. Lett. {\bf 122}, 042202 (2023).
\bibitem{SCES2023} Y. Funami and K. Aoyama, New Phys.: Sae Mulli {\bf 73}, 1086 (2023).
\bibitem{Nikitin2023} M. V. Nikitin, V. Ya. Pokrovskii, D. A. Kai, and S. G. Zybtsev, JETP Lett. {\bf 118}, 861 (2023).
\end{thebibliography}

\begin{thebibliography}{80}
\bibitem{SMatsukawa_Takayama} H. Matsukawa and H. Takayama, J. Phys. Soc. Jpn. {\bf 56}, 1507 (1987); H. Matsukawa, {\it ibid}. {\bf 56}, 1522 (1987); {\bf 57}, 3463 (1988).
\bibitem{SPRBL2023} Y. Funami and K. Aoyama, Phys. Rev. B {\bf 108}, L100508 (2023).
\end{thebibliography}
\end{document}